\documentclass[conference,10pt]{IEEEtran} 
\IEEEoverridecommandlockouts
\makeatletter
\def\ps@headings{
\def\@oddhead{\mbox{}\scriptsize\rightmark \hfil \thepage}
\def\@evenhead{\scriptsize\thepage \hfil \leftmark\mbox{}}
\def\@oddfoot{}
\def\@evenfoot{}}
\makeatother
\usepackage{colortbl}
\usepackage{graphicx}
\usepackage{tabularx}
\usepackage{times}
\usepackage{amsmath}
\usepackage{amsthm}
\usepackage{amssymb}
\usepackage{cite}
\usepackage{epsfig}
\usepackage{epstopdf}
\usepackage{subfigure}
\usepackage[ruled]{algorithm2e}
\usepackage{mathtools}
\usepackage{float}
\usepackage{setspace}
\usepackage{mathtools}
\usepackage[ruled]{algorithm2e}
\usepackage{algorithmic}
\usepackage[justification=centering]{caption}
\usepackage{graphicx}
\usepackage{subcaption}
\DeclareGraphicsExtensions{.eps}
\begin{document}
\title{\huge{ Emergency Caching: Coded Caching-based Reliable Map Transmission in Emergency Networks}\vspace{-0.5cm}}
\author{\IEEEauthorblockN{Zeyu Tian\IEEEauthorrefmark{1},
Lianming Xu\IEEEauthorrefmark{2},  Liang Li\IEEEauthorrefmark{3}, Li Wang\IEEEauthorrefmark{1}, and Aiguo Fei\IEEEauthorrefmark{1}
\vspace{-0.2cm}
}\\
\small\centerline{\IEEEauthorrefmark{1}School of Computer Science, Beijing University of Posts and Telecommunications, Beijing, China}\\\IEEEauthorrefmark{2}
School of Electronic Engineering, Beijing University of Posts and Telecommunications, Beijing, China\\\IEEEauthorrefmark{3}Frontier Research Center, Peng Cheng Laboratory, Shenzhen, China\\
Email:{\{tianzeyu, xulianming\}@bupt.edu.cn, lil03@pcl.ac.cn, \{liwang, aiguofei\}@bupt.edu.cn }\\
\vspace{-1cm}

        }\vspace{-0.5cm}
\maketitle
\renewcommand{\thefootnote}{}
\footnote{\indent This work was supported in part by the  National Natural Science Foundation of China under grants U2066201, 62171054, and 62201071,  in part by the Fundamental Research Funds for the Central Universities under Grant No.24820232023YQTD01, and in part by the Interdisciplinary Team Project Funds for the Double First-Class Construction Discipline under Grant No. 2023SYLTD06. (Corresponding author: Li Wang.)}
\begin{abstract}
Many rescue missions demand effective perception and real-time decision making, which highly rely on effective data collection and processing. In this study, we propose a three-layer architecture of emergency caching networks focusing on data collection and reliable transmission, by leveraging efficient perception and edge caching technologies. Based on this architecture, we propose a disaster map collection framework that integrates coded caching technologies. Our framework strategically caches coded fragments of maps across unmanned aerial vehicles (UAVs), fostering collaborative uploading for augmented transmission reliability. Additionally, we establish a comprehensive probability model to assess the effective recovery area of disaster maps. Towards the goal of utility maximization, we propose a deep reinforcement learning (DRL) based algorithm that jointly makes decisions about cooperative UAVs selection, bandwidth allocation and coded caching parameter adjustment, accommodating the real-time map updates in a dynamic disaster situation. Our proposed scheme is more effective than the non-coding caching scheme, as validated by simulation.
\end{abstract}
\thispagestyle{empty}

\section{Introduction}
In recent years, the surge in catastrophic events has significantly impacted the lives of individuals worldwide. Central to the efficacy of the disaster response process is the timely detection of critical circumstances and the continuous updating of situational information. However, in disaster scenarios, on the one hand, existing wireless network infrastructures may be severely damaged such that Reduces the reliability of emergency communications \cite{yingji}. On the other hand, with the diverse range of data applications such as high-definition (HD) maps\cite{livemap}, disaster situation and emergency services, which introduces a heavy burden to the wireless network\cite{disaster}.\par
Therefore, substantial research  works are needed to develop for providing practical and effective emergency situation transmission services. The disaster situation maps encapsulate dynamic information, necessitating regular updates to ensure their accuracy. Previous research has delved into the transmission of situation maps. In \cite{1}, CarMap can realize real-time updates on the feature-represented map by excluding transient information from map processing, but it transmits a simplified version of the map. In a separate study \cite{jiqunHD}, the authors introduce a method for transmitting High-Definition maps and subsequently analyze the associated offloading delay. However, due to inclusion of extensive dynamic information with situation maps, transmitting large files significantly reduces transmission reliability. It's worth noting that neither of these studies addresses reliable transmission and update of large data volumes under bandwidth constraints. Coded caching technology can provide redundancy by encoding the files into miniature fragments and storing them in different nodes to increase transmission reliability\cite{6--}. 
In order to improve the efficiency of content sharing between devices, the collaborative coding caching technique is applied in \cite{d2w}. Works in \cite{521} proposed a file partitioning and grouping scheme in a distributed coded cache system. This scheme designed a static single-server system for heterogeneous file transfer, but it was not suitable for dynamic change and actual channel distribution. None of the above works considers the effect of user mobility. In \cite{12--}, the concept of caching coded content directly onto mobile devices is explored, yet variations in different coding parameters are not addressed. However, the creation of the disaster map demands a holistic approach that encompasses both the dynamic content updates and the transmission of multi-scale content within the constraints of limited spectrum resources and node mobility. \par
To tackle the challenges above, in this work, we propose an emergency caching network architecture, which can  facilitate the disaster information reliable transmission. In particular, a coded caching-enabled disaster map transmission framework is designed based on emergency caching networks. Subsequently, we establish the ground vehicle's (GV) probability of successful map recovery through coded fragments aggregation, forming the basis for defining the coupling between reliable transmission and effective coverage. We propose a Deep Reinforcement Learning (DRL)-based algorithm to facilitate real-time maps update by concurrently optimizing cooperative unmanned aerial vehicles (UAVs) selection, bandwidth allocation, and coding parameter adjustments. Lastly, we validate the superior performance of the proposed scheme through an extensive series of simulation experiments.
\vspace{-0.1cm}

\begin{figure*}[t]
  \centering
  \subfigure[Emergency caching networks.] {\label{com_dis_2_hit}  \includegraphics[width=0.3\textwidth]{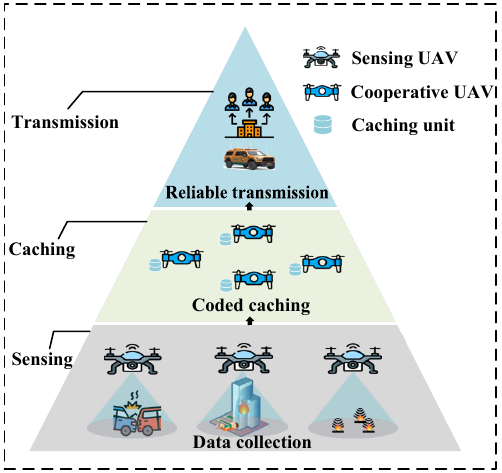}}
  \subfigure[The diagram of emergency caching networks.] {\label{com_dis_2_utility}
 \includegraphics[width=0.65\textwidth]{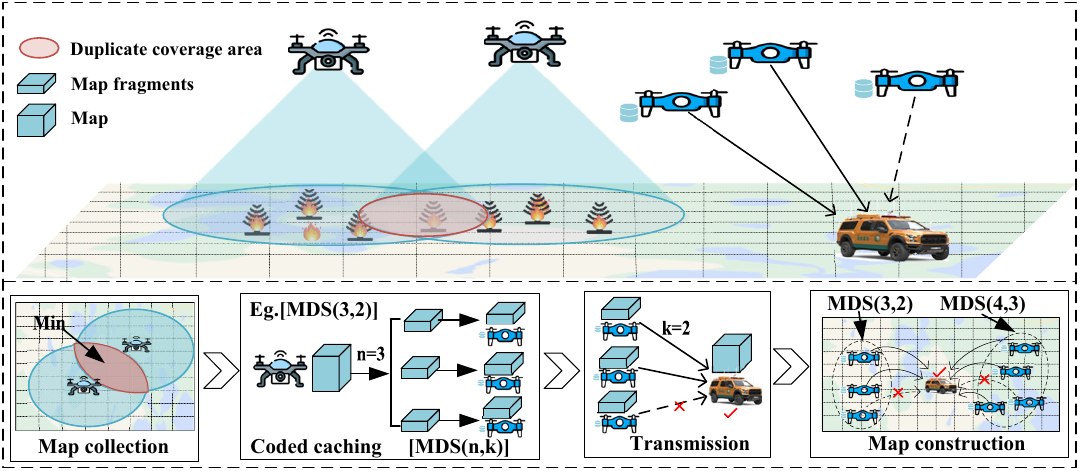}}
  \caption{{The hierarchical architecture of emergency caching networks.}}\label{fig:background}
  \vspace{-0.7cm}
\end{figure*}
\section{Architecture of Emergency Caching Networks}
To effectively carry out critical tasks such as sensing, document dissemination, and situational map construction in emergency rescue operations, a three-layer caching network architecture is proposed, as shown in Fig. \ref{com_dis_2_hit}. The three layers of the network architecture are introduced in the following.\par
$\bullet$ \textbf{\emph{Sensing layer:}} 
 Different terminal devices such as radar and cameras play a crucial role in the perception and acquisition. Due to their flexibility and continuously improving capabilities, UAVs will assist in various tasks, for example, data collection or acting as aerial base stations.\par
$\bullet$ \textbf{\emph{Caching layer:}} Within the emergency caching network, cache nodes are typically set as UAVs. These nodes possess excellent storage capabilities, which further enhance their ability to perform caching tasks. Multiple cache UAVs can collaborate to quickly establish the emergency caching networks and enhance its network capacity.\par
$\bullet$ \textbf{\emph{Transmission layer:}} The transmission layer receives and processes large data collected from the field, providing valuable insights for on-site emergency command. However, the unreliable nature of the field links poses a challenge to the reliable transmission of data. It requires the coordination of the three-layer network architecture to ensure the reliable and timely transmission of data.\par
We will provide a tangible illustration of the benefits of our proposed architecture by considering an example of transmitting disaster maps through the emergency caching network.
\section{Case Study: Coded caching aided reliable disaster maps transmission}
\subsection{System Model}
We consider a UAV-assisted emergency caching network, as shown in Fig. \ref{com_dis_2_utility}, where several geographically dispersed UAVs capture their surrounding scene with the equipped camera and transmit the images to the GV for disaster map construction. Assuming there are two types of UAVs, namely sensing UAV (SU) and cooperative UAV (CU). The former generates images, while the latter cooperatively assists in the transmission stage. We index the SUs by $\mathcal{I}=\left\{ {1,...,i,...,I} \right\}$. The target area is mapped to a rasterized geospatial index structure and divided into $C$ fine-grained grids, denoted by ${\cal C}=\left\{{1,...,c,...,C}\right\}$. The sensing area of the $i$-th SU is denoted by
${{\rm{{\cal L}}}_i} = \left\{ {1,...,l,...,L} \right\} \subset {\rm{{\cal C}}}$, where ${{\rm{{\cal L}}}_{{\rm{  }}i}}$ is the set of areas with coverage area unit indices as elements. \par
We choose the MDS-coded method with the coded caching parameter set to ${(n,k)}$ \cite{MDS}, where $n$ denotes the number of redundant coded fragments and $k$ indicates how many fragments are needed to recover the file. Suppose the disaster map collected by the $i$-th SU is recorded as ${f_{i}}$, the file size is recorded as ${{Z}_{i}},\forall i\in \mathcal{I}$. The redundantly coded fragments are stored at nearby CUs, and the size of each fragment is ${{{\alpha }_{i}}={{Z}_{i}}/{{k}_{i}}}$, where ${k}_{i}$ is the number of fragments required to restore the map file perceived by the $i$-th SU. We assume that SUs can successfully share the fragments with CUs, and there are enough CUs to assist SUs.

\subsubsection{Transmission Model}
We apply orthogonal frequency division multiple access (OFDMA) as the accessing scheme for different communication links. We assume that the Air-to-Ground channel obeys the Rician fading model. We define the set of CUs that assists with transmitting file $f_i$ as ${\mathcal{U}_{i}}=\left\{{U\!AV_{1_i},...,U\!AV_{e_i},...,U\!AV_{E_i}} \right\}$, where ${U\!AV}\!_{e_i}$ represents the $e$-th CU assisting in the transfer of file $f_i$. Then, the GV receives the signal sent by ${U\!AV}\!_{e_i}$ can be expressed as

\vspace{-23pt}
\begin{center}\begin{equation}y_{e_i,g}^{{}}=\sqrt{{{p}}}h_{e_i,g}x_{e_i,g}+{{n}_{n}},\end{equation}
\end{center}
\vspace{-5pt}
where ${p}$ and $x_{e_i,g}$ represent the transmit power and signal from $U\!AV_{e_i}$, respectively. $h_{e_i,g}=\sqrt{\beta_{e_i,g}}g_{e_i,g}$ represents the channel gain, and $\beta_{e_i,g}=\beta_{0}(d_{e_i,g})^{-\alpha}$ is the large-scale average channel power gain, $d_{e_i,g}$ is the distance between the $U\!AV_{e_i}$ and GV. $g_{e_i,g}$ is the small-scale fading coefficient. In addition, ${{n}_{n}}\sim \mathcal{C}\mathcal{N}\left( 0,\sigma _{0}^{2} \right)$ is additive white Gaussian noise, and ${{\sigma} _{0}^{2}}$ is the noise power. 
Therefore, the data rate can be expressed as ${{{R}_{e_i,g}}={{\log }_{2}}\left( 1+{{\gamma }_{e_i,g}} \right)}$,
where ${\gamma _{e_i,g}^{{}}={{p}}{{\left| h_{e_i,g}^{{}} \right|}^{2}}/\sigma _{0}^{2}={{p}}{{\mu }_{e_i,g}}}$ is the corresponding signal-to-noise ratio (SNR).
The recovery probability of the disaster map is related to the successful transmission probability (STP) between the GV and the CU. $B_i$ is defined as the bandwidth allocated to the SU $i$ as well as the CU for its transmission. Define ${{\eta }_{e_i,g,{{f}_{i}}}}$ as the STP that a coded fragment of file $f_i$ was sent by the ${U\!AV}\!_{e_i}$ to the GV. It is calculated as
\vspace{-0.8cm}
\begin{center}\begin{equation}\label{gs2}
{{{\eta }_{e_i,g,{{f}_{i}}}}=\Pr \left\{ {{T}_{e_i,g}}\ge ({{{\alpha }_{i}}}/({{{({B}_{i}/k_i)}}{{R}_{e_i,g}}}) \right\}},\end{equation}
\end{center}
where ${{T}_{e_i,g}}$ is the contact time of the ${U\!AV}_{e_i}$ and the GV, ${B_i/k_i}$ is the bandwidth allocated to the ${U\!AV}_{e_i}$.\par
We apply the Rician fading model with the mean channel power gain ${\overline{\mu }_{e_i,g}}$ and the Rician factor $\chi$. Then the Cumulative Density Function (CDF) of ${{\mu }_{e_i,g}}$ is expressed as \cite{14--}
\begin{align}
{{F}_{{{\mu }_{e_i,g}}}}\left( x \right)=1-{{e}^{-\chi }}\sum\limits_{{{m}_{0}}=0}^{\infty }{\sum\limits_{l=0}^{{{m}_{0}}}{\frac{{{\chi}^{{{m}_{0}}}}}{{{m}_{0}}!l!}{{x}^{l}}{{\left( {{\zeta }_{e_i,g}} \right)}^{l}}}}{{e}^{-{{\zeta }_{e_i.g}}x}},
\end{align}
where ${\zeta _{e_i,g}=\left( \chi+1 \right){{\left( \overline{\mu }_{e_i,g}^{{}} \right)}^{-1}}}$. Thus, the CDF of  ${{R}_{e_i,g}}$ is derived as 
\begin{small}
    \begin{align}
{{F}_{{{R}_{e_i,g}}}}\left( r \right)&=\Pr \left\{ {{R}_{e_i,g}}\le r \right\} ={{F}_{{{\mu}_{e_i,g}}}}\left(({{{2}^{r}}-1})/{{{p}}} \right).
\end{align}
\end{small}
We assume that the contact time ${{T}_{e_i,g}}$ between the ${U\!AV}\!_{e_i}$ and the GV follows an exponential distribution with mean $\tau $, so ${{{\widehat{T}}_{e_i,g}}={{T}_{e_i,g}}/\tau}$ obeys an exponential distribution with a unit mean\cite{contact time}. Therefore, Eq. (\ref{gs2}) can be derived as
\begin{small}
\begin{align}
\setlength{\abovedisplayskip}{-3pt}
\setlength{\belowdisplayskip}{3pt}
 {{\eta }_{e_i,g,{f_i}}}&=\int_{0}^{\infty }{\left( 1-{{F}_{{{R}_{e_i,g}}}}\left( \frac{{{\alpha }_{i}}}{\tau {{({B}_{i}/k_i)}}t} \right) \right)}{{e}^{-t}}dt, 
\end{align}
\end{small}

where ${{{F}_{{{R}_{e_i,g}}}\left( r \right)}={{{F}_{{{\mu}_{e_i,g}}}}\left({{({2}^{r}}-1)} / {p} \right)}}$.

\subsubsection{Problem Formulation}
It is crucial to maximize the recovery area of the disaster map while balancing the system performance. Therefore, we propose a new metric to carve out the effective coverage area of the map. Assuming that the observation period is divided into $T$ time slots, represented as $\mathcal{T}=\left\{ 1, \ldots , t, \ldots , T \right\}$. We define ${{Pr_{c}{(t)}}}$ as the recovery probability of the unit area ${c}$ at time slot $t$, expressed as

\vspace{-15pt}
\begin{small}
\begin{align}
\label{gl}
{{\Pr }_{c}{(t)}}=1-\prod\limits_{i:\text{ c}\in {{\mathsf{\mathcal{L}}}_{i}}}{\left( 1-{{x}_{i}{(t)}}{{P}_{c,{{f}_{i}}}{(t)}} \prod\limits_{e_i\in \mathcal{E}_{{{f}_{i}}}^{\text{ }opt}}{{{\eta }_{e_i,g,{{f}_{i}}}{(t)}}} \right)},
\end{align}
\end{small}
where ${{P}_{c,{{f}_{i}}}{(t)}}$ represents the recovery probability of the $i$-th SU covering the unit area $c$, and ${{P}_{c,f_i}{(t)}={{P}_{{{f}_{i}}}{(t)}}}$. The specific derivation of ${P}_{c,f_i}{(t)}$ is shown in Eq. (\ref{p_succ2}). ${{x}_{i}{(t)}}\in \left\{ 0,1 \right\}$, $\forall i\in \mathcal{I}$, indicates whether to schedule the ${i}$-th SU perform the mapping task. Eq. (\ref{gl}) represents the probability that at least one SU will be able to sense region $c$ and recover successfully. The smaller the overlapping collection area, the higher the efficiency. Thus, the effective recovery area can be expressed as ${C_{sum}(t)= {\textstyle \sum_{c\in \cal C }Pr_c(t)A}}$, where ${A}$ represents the size of the area $c$.\par
 We aim to maximize the disaster map effective recovery area over a finite time horizon of $T$ time slots. For each time slot, we define $\boldsymbol{x}(t)=(x_i)_{1\times\left|\mathcal{I}\right|}$ as the scheduling variable of the SU, $\boldsymbol{b}(t)=(B_i)_{1\times\left|\mathcal{I}\right|}$ as the bandwidth allocation variable, $\boldsymbol{k}(t)=(k_i)_{1\times\left|\mathcal{I}\right|}$ as the coded caching parameters adjustment variable.  The problem can be formulated as follows

\vspace{-0.5cm}
\begin{small}
\begin{subequations}\label{obj1}
\begin{align}
\mathcal{P}:\ &
\scalebox{1.0}{$\underset{\left\{ \boldsymbol{x}(t),\boldsymbol{b}\left( t \right),\boldsymbol{k}(t) \right\}}{\mathop{\max }}\,\text{   }\frac{1}{T}\text{ }\sum\limits_{t=1}^{T}{{\textstyle \sum\limits_{c\in \cal C }Pr_c(t)}}$}\\
\emph{\text{s.t.}} \quad &\scalebox{0.95}{$\sum\limits_{e_i\in \mathcal{E}_{{{f}_{i}}}^{opt}}{{{\pi }_{e_i,g,{{f}_{i}}}}}\left( t \right)={{k}_{i}},\forall i\in \mathcal{I},t\in \mathcal{T}$},\\
&\scalebox{1.0}{${{x}_{i}}(t)\in \left\{ 0,1 \right\},\forall i\in \mathcal{I},t\in \mathcal{T}$},\\
&\scalebox{1.0}{$\sum\limits_{1}^{I}{{{x}_{i}}\left( t \right){{B}_{i}}\left( t \right)}\le {{B}_{total}},\forall i\in \mathcal{I}, t\in \mathcal{T}$},\\
&\scalebox{1.0}{${{\gamma }_{e_i,g,{{f}_{i}}}}{(t)}\ge {{\gamma }_{0}},\forall i\in \mathcal{I},e_i\in \mathcal{E}_{{{f}_{i}}}^{opt},t\in \mathcal{T}$}.
\end{align}
\end{subequations}
\end{small}

Constraint (7b) indicates that ${k_i}$ fragments need to be obtained when the GV restores the map of area ${{\rm{{\cal L}}}_i}$. Constraint (7c) is the scheduling of SUs. Constraint (7d) is the total bandwidth constraint. Constraint (7e) is the SNR threshold. In problem (7), the disaster map is dynamic update over time. Dynamic sequential decisions, such as those pertaining to bandwidth allocation and UAVs scheduling, wield significant influence on future outcomes. The process exhibit Markovian properties. Moreover, the absence of dynamic adaptation in traditional schemes arises from dynamic changes in channel parameters and the time-dependent adjustment of coding parameters. In this work, we propose a DRL-based algorithm to solve this problem.
\vspace{-0.3cm}

\subsection{Proposed Algorithm And Analysis}

\begin{algorithm}[t]
\small
\renewcommand{\algorithmicrequire}{\textbf{Inputs:}}
\renewcommand{\algorithmicensure}{\textbf{Outputs:}}
\caption{SACRL Algorithm}
\label{alg-matching}
\begin{algorithmic}[1]
\REQUIRE{Randomly initialize Q-network and target Q-network with weights ${\theta}$ and $\theta^{’}$}. Initialize replay memories ${D}$.
\ENSURE Q-network parameters. \\
\FOR {epoch = $1$ $to$ $N$}
\STATE Initialize state ${\boldsymbol{s}_1=\left[ \boldsymbol{p}_1,\boldsymbol{f}_1 \right]}$.
\FOR {t = $1$ $to$ $T$}
\STATE With probability $\varepsilon$ select a random action, otherwise select $\boldsymbol{a}_t=\arg \max _{a} Q\left(\boldsymbol{s}_t, \boldsymbol{a}_t; \theta\right)$.
\STATE Execute action $\boldsymbol{a}_t$ and observe $r_t$ and $\boldsymbol{s}_{t+1}$.
\STATE Store the experience $(\boldsymbol{s}_t, \boldsymbol{a}_t, r_t, \boldsymbol{s}_{t+1})$ into $D$.
\STATE Get a random mini-batch of $M$ samples from $D$.
\STATE Set $y_{i}=r_{i}+\gamma \max _{a^{\prime}} Q^{\prime}\left(s_{i+1}, a^{\prime} ; \theta^{\prime}\right)$.
\STATE Perform a gradient descent step on (\ref{dq}) with respect to the  network parameters $\theta$.\\
Every $C$ steps reset $\theta^{'}=\theta$.
\ENDFOR
\ENDFOR
\end{algorithmic}
\end{algorithm}

   
In this section, we analyze the selection of collaborative UAVs. Then, we propose a DRL-based joint cooperative UAVs selection, bandwidth allocation, and coding parameters algorithm (SACRL) to solve the original problem (\ref{obj1}).
\subsubsection{Cooperative UAV Selection scheme}
 Define the index set of CUs that store file $f_i$ fragments as ${{{\varepsilon}}_{i}}=\left\{ 1_i,...,e_i,...,E_i \right\}$, where ${\left| {{{\varepsilon}}_{i}} \right|=E_i}$. When the SNR of any UAV is greater than the minimum SNR requirement ${{\gamma }_{0}}$, it can be used as a CU. The set of CUs is denoted by
\vspace{-0.9cm}
\begin{center}\begin{equation}
{{\mathcal{E}}_{{{f}_{i}}}}=\left\{ e_i|{{\gamma }_{e_i,g,{{f}_{i}}}}\ge {{\gamma }_{0}},\forall e_i\in {{{\varepsilon}}_{i}} \right\}, \end{equation} 
\end{center}
where ${{\gamma }_{e_i,g,{{f}_{i}}}}$ is the SNR of the ${U\!AV}\!_{e_i}$ that delivers the coded fragment of file $f_i$ to the GV.\par
Assuming that ${m}$ CUs in ${\mathcal{E}}_{{{f}_{i}}}$ can successfully transmit the map fragments, there are ${{C}_{\left| {{\mathcal{E}}_{{{f}_{i}}}} \right|}^{m}={\left( \left| {{\mathcal{E}}_{{{f}_{i}}}} \right| \right)!}/{\left( \left( m \right)!\left( \left| {{\mathcal{E}}_{{{f}_{i}}}} \right|-m \right)! \right)}}$ possible combinations. Define $\mathcal{C}{{\mathcal{N}}_{L}}\left( m \right)=\left\{ {{e}_{i1}}, \ldots ,{{e}_{im}} \right\}\subset {{\mathsf{\mathcal{E}}}_{{{f}_{i}}}}$ to represent all possible choices of CUs, with cardinality ${\left| \mathcal{C}{{\mathcal{N}}_{L}}\left( m \right) \right|=m}$, where ${L\in \left\{ 1, \cdots , C_{\left| {{\mathsf{\mathcal{E}}}_{{{f}_{i}}}} \right|}^{m} \right\}}$. Meanwhile, the complement of $\mathcal{C}{{\mathcal{N}}_{L}}\left( m \right)$ can be expressed as ${\overline{\mathcal{C}{{\mathcal{N}}_{L}}}\left( m \right)=\left\{ {{
e}_{i(m+1)}}, \ldots , {{e}_{\left| {{\mathcal{E}}_{{{f}_{i}}}} \right|}} \right\}}$. Therefore, for any possible $\mathcal{C}{{\mathcal{N}}_{L}}\left( m \right)$, the probability that a total of ${m}$ links successfully transmit can be expressed as ${\prod\nolimits_{e_i\in \mathcal{C}{{\mathcal{N}}_{L}}\left( m \right)}{{{\eta }_{e_i,g,{{f}_{i}}}}}\prod\nolimits_{e_i\in {{\overline{\mathcal{C}\mathcal{N}}}_{L}}\left( m \right)}{\left( 1-{{\eta }_{e_i,g,{{f}_{i}}}} \right)}}$.\par
The file ${{f}_{i}}$ can only restore when $m\ge {{k}_{i}}$. Therefore, the probability that the GV can successfully recover the disaster map collected by the $i$-th SU is expressed as

\vspace{-8pt}
{\setlength\abovedisplayskip{3pt}
 \setlength\belowdisplayskip{1pt}
\begin{small}
\begin{align}\label{p_succ2}
&{P_{{f}_{i}}} = \begin{array}{l}
\sum\limits_{m={{k}_{i}}}^{\left|{{\mathcal{E}}_{{{f}_{i}}}} \right|}\!{\sum\limits_{L=1}^{C_{\left|{{\mathcal{E}}_{{{f}_{i}}}} \right|}^{m}}\!{\left( \prod\limits_{e_i\in\mathcal{C}{{\mathcal{N}}_{L}}\left(m\right)}\!{\!{{\eta}_{e_i,g,{{f}_{i}}}}}\!\!\prod\limits_{e_i\in \overline{\mathcal{C}{{\mathcal{N}}_{L}}}\left(m\right)}\!\!\!\!{\left( 1-{{\eta }_{e_i,g,{{f}_{i}}}} \right)} \right)}}.
\end{array} 
\end{align}
\end{small}}\par
To maximize the probability that the GV restores the file ${f_i}$. It is necessary to select ${k_i}$ CUs with the largest STP from the set ${{{\mathcal{E}}_{{{f}_{i}}}}}$ to transfer the coded fragment. ${{{\pi }_{e_i,g,{{f}_{i}}}}}$ is a binary variable. If the ${U\!AV}\!_{e_i}$ delivers a coded fragment of file $f_i$ to the GV, ${\pi_{e_i,g,f_i}}=1$, otherwise ${\pi_{e_i,g,f_i}}=0$. Therefore, 
the optimal set of CUs that assist in transmitting coded fragments of file ${f_i}$ to the GV is $\mathcal{E}_{{{f}_{i}}}^{opt}=\left\{ e_i|{{\pi}_{e_i,g,{{f}_{i}}}}=1,\forall e_i\in {{\mathcal{E}}_{{{f}_{i}}}} \right\}$.
\subsubsection{SACRL algorithm}
\begin{figure}[t]
\centering
\includegraphics[width = 0.49\textwidth]{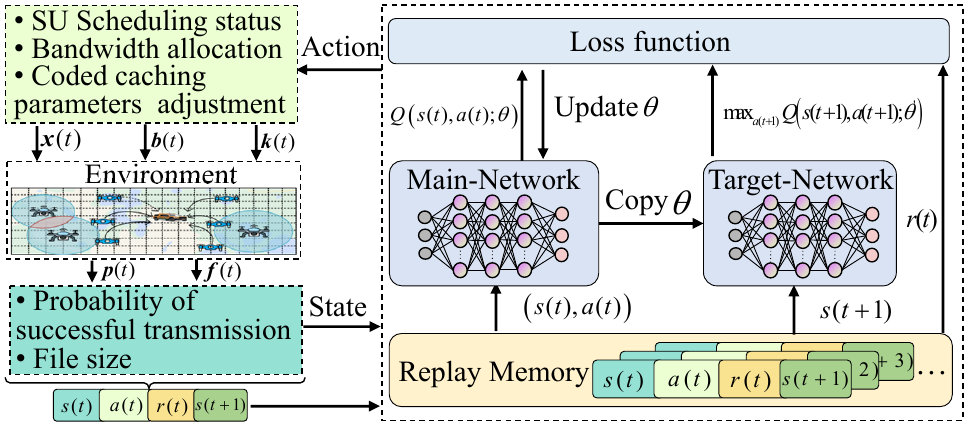}
\caption{The framework of SACRL algorithm.}
\label{fig:cluster}
\vspace{-25pt}
\end{figure}
 In Fig. 2, GV makes decisions by interacting with the environment over time \cite{dqn}. The three key elements of the MDP model are defined as follows.

$\bullet$ \textbf{\emph{State:}} The state at the $t$-th time slot is expressed as 
\vspace{-3pt}
\begin{align}
\boldsymbol{s}\left( t \right)=\left[ \boldsymbol{p}(t),\boldsymbol{f}(t) \right],
\end{align}
where $\boldsymbol{p}(t)=\left[ {{P}_{{f}_1}}\left( t \right),...,{{P}_{{f}_i}}\left( t \right),...,{{P}_{{f}_I}}\left( t \right) \right]$ represents the probability matrix that the file was successfully transmitted by the CU to the GV by coded caching, $\boldsymbol{f}(t)=\left[ {{\alpha}_{1}}(t),...,{{\alpha}_{i}}(t),...,{{\alpha}_{I}}(t) \right]$ represents the file size of the disaster map constructed by the UAV in the $t$-th time slot.

$\bullet$ \textbf{\emph{Action:}} The action of the $t$-th time slot is defined as the cooperative UAVs selection, the bandwidth allocation, and the adjustment strategy of coding parameters. Here, we discretize the bandwidth resource into $h$ optional values, denoted as ${\mathsf{\mathcal{B}}=\left[ {{b}_{1}},...,{{b}_{h}} \right]}$. Therefore, the action can be expressed as
\vspace{-3pt}
\begin{align}
\boldsymbol{a}(t)=\left[ \boldsymbol{x}\left( t \right),\boldsymbol{b}\left( t \right),\boldsymbol{k}\left( t \right) \right],
\end{align}
where $\boldsymbol{x}\left( t \right)=\left[ {{x}_{1}}(t),...,{{x}_{i}}(t),...,{{x}_{I}}(t) \right]$, where ${{x}_{i}}(t)=\left\{ 1,0 \right\}$, represents the scheduling status of the sensing UAV for each time slot. ${{{\boldsymbol{b}}\left( t \right)}=\left[ {{B}_{1}}\left( t \right),...,{{B}_{i}}\left( t \right),...,{{B}_{I}}\left( t \right) \right]}$, where ${B}_{i}\in\mathsf{\mathcal{B}},\forall i\in \mathcal{I}$ represents the bandwidth resource allocated to the UAVs for cooperative transmission of the file $f_i$. $\boldsymbol{k}(t)=\left[ {{k}_{1}}\left( t \right),...,{{k}_{i}}\left(t\right),...,{{k}_{I}}\left( t \right) \right]$ represents the adjustment of coded caching parameters in $t$-th time slot.

$\bullet$ \textbf{\emph{Reward:}} The reward at the $t$-th time slot is defined as a linear function of the disaster map effective recovery area, denoted as ${r(t)=\lambda ( {\textstyle \sum_{c\in \cal C}P_{r_c}(t)A } )}$, where ${\lambda}$ is a constant\cite{drl}. The goal of learning is to maximize the gain defined as the expected discounted future reward, it can be expressed as ${R(t)=\mathbb{E} \left [  {\textstyle \sum_{{t^{'}}=t}^{T}}\gamma^{t^{'}-t}r(t^{'}) \right ] }$, where $T$ and ${0<{\gamma}<1}$ are the terminal step of each episode and discount factor that represents the impact of future reward, respectively. 

The solution is to find the optimal policy that maximizes the expected return. Deep Q-network has two neural networks. We use a neural network ${Q(s,a;\theta)}$, called a prediction network, as a function approximator to estimate the action value function, ${\theta}$ is the weight of the neural network. Furthermore, another neural network ${Q^{'}(s,a;\theta^{'})}$ with the same structure, called the target network, is also used to estimate the target value in (\ref{dq}). However, its weights ${\theta^{'}}$ are copied from ${\theta}$ every fixed number of iterations $n$ instead of every training epoch. The loss function at each step is expressed as 
\vspace{-5pt}
\begin{align}
\begin{small}
    \label{dq}
L(\theta)=\mathbb{E}\left [ ({y_j-Q({\mathbf{s}_j},{\mathbf{a}_j};\theta ))}^{2}  \right ] ,
\end{small}
\end{align}
where ${{y}_{j}}={{r}_{j}}+\gamma {{\max }_{{{a}^{'}}}}{{Q}^{'}}({{s}_{j+1}},{{a}^{'}};{{\theta }^{'}})$. The main solution process of the problem can be summarized as \textbf{Algorithm 1}.
\vspace{-3pt}
\subsection{Simulation Results and Analysis}
\begin{table}[t]
  \caption{Simulation Parameters}\label{tab_sim}
  \small
 \begin{center}
    \begin{tabular}{l|l}
    \hline\hline
       \bf{Parameter} &\bf{Value}\\
      \hline 
         Target area & $1000m\times1000m$\\ 
         Total bandwidth $B_{total}$ & $10MHz$\\
         Memory size $M$& 2000\\
         Learning rate $\varphi$ & 0.001 \\
         Noise power ${\sigma_0^2}$ & $-90dBm$\\
         Transmit power $p$ & $0.15w$\\
         SNR threshold $\gamma_0$ &  $27dB$\\
         Rician factor $\chi $ & 3 \\
         Pathloss exponent $\alpha$ & 4 \\
        The number of sensing UAVs & 2\\
        The number of ancillary UAVs & 16\\
        The flight altitude of UAVs &$100m$\\
    \hline\hline
    \end{tabular}
    \end{center}
    \vspace{-10pt}
\end{table}

\begin{figure*}[t]
	\setlength{\abovecaptionskip}{-1pt}
	\setlength{\belowcaptionskip}{-1pt}
	\centering
	\begin{minipage}{0.33\linewidth}
		\centering
        \subfigure[Coverage apothem.] {\label{fig:PANDGAMA}
    \includegraphics[width=0.45\textwidth]{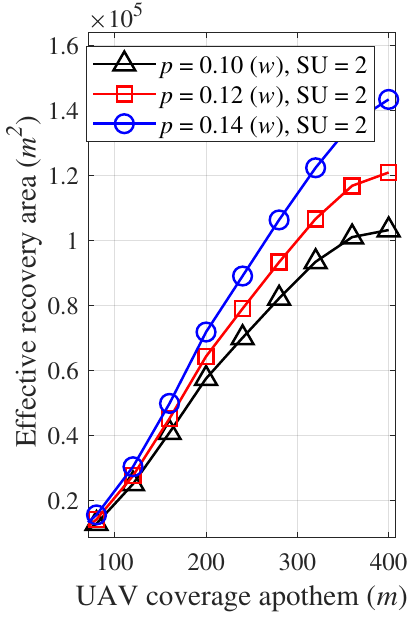}}
        \subfigure[Content size.] {\label{fig:PANDGAMA2}
    \includegraphics[width=0.45\textwidth]{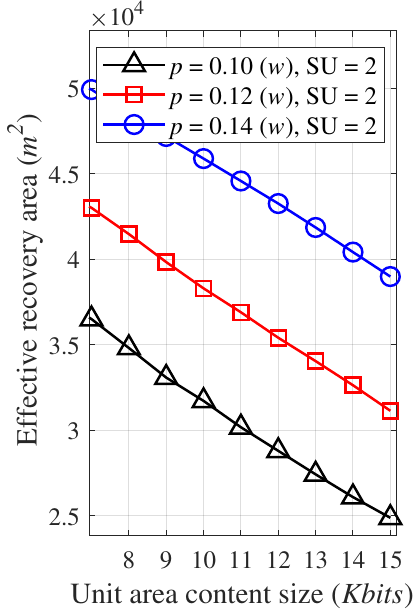}}
        \caption{Effects of system parameters.}\label{MC_1}
	\end{minipage}%
	\begin{minipage}{0.33\linewidth}
		\centering
        \subfigure[Transmission power.] {\label{fig:sanzhong2}
    \includegraphics[width=0.45\textwidth]{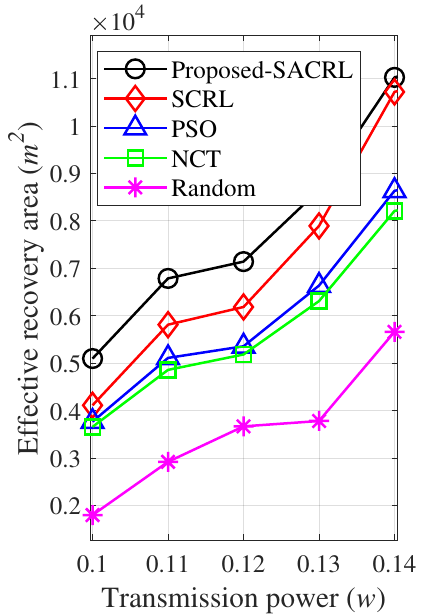}}
   \subfigure[Content size.] {\label{fig:sanzhong1}
    \includegraphics[width=0.45\textwidth]{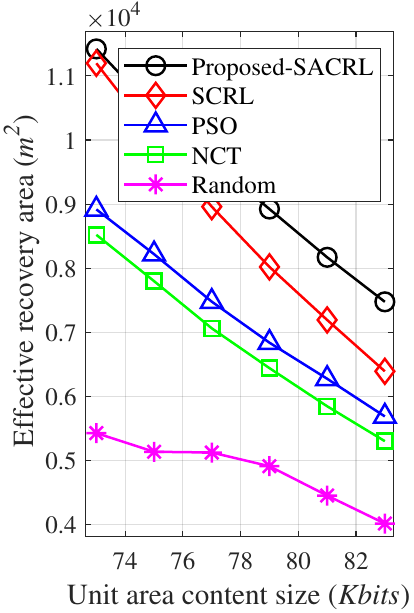}} 
        \caption{Effects of different methods.}\label{MC_1}
	\end{minipage}
	\begin{minipage}{0.33\linewidth}
		\centering
    \includegraphics[width=0.85\textwidth]{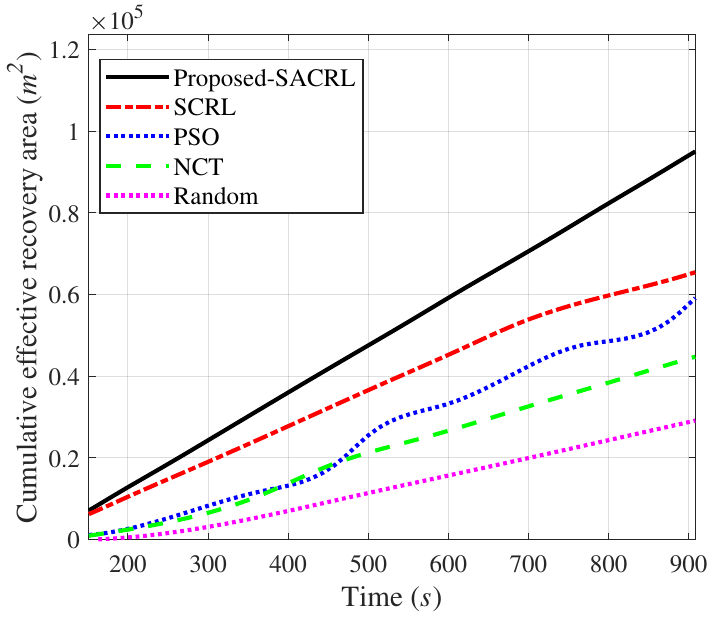}
		\caption{Recovery area over time.}\label{fig:time}
	\end{minipage}
    \vspace{-18pt}
\end{figure*}

\vspace{-3pt}
In this section, we perform simulations to verify the effectiveness of the proposed algorithm. Some fixed parameters are shown in the TABLE \ref{tab_sim}. The UAVs perform the construction task of the disaster map at a fixed altitude in a specific target area, and move randomly within the target area. The GV is located at the center of the target area. The transmission bandwidth strategy of the GV is divided into three levels: ${\mathcal{B}=\left \{ 2MHz, 6MHz \right \} }$. The coded caching parameter adjustment strategy is divided into two levels: ${k_i=\left \{ 1, 2 ,3 \right \} }$. In our proposed algorithm, we use ReLu as the activation function. We compare our method with the following benchmarks:
\begin{itemize}
    \item \textbf{\emph{DRL-based UAV selection algorithm(SCRL)}}: This method jointly optimizes cooperative UAVs and coding parameters selection with average bandwidth allocation.
    \item \textbf{\emph{Particle Swarm Optimization algorithm(PSO)}}: PSO is used to optimize cooperative UAVs, bandwidth, and coding parameters selection \cite{pso}.
    \item  \textbf{\emph{Non-coded caching transmission algorithm(NCT)}}: This method uses a non-coded caching method for file transmission \cite{noncode}. In order to ensure fairness, bandwidth and cooperative UAVs selection are jointly optimized.
    \item \textbf{\emph{Random algorithm}}: Cooperative UAVs, bandwidth level, coding parameters are randomly selected.
\end{itemize} \par  
\vspace{-3pt}
Fig. \ref{fig:PANDGAMA} shows the comparison of the effective recovery area as the UAV coverage sensing radius increases. In the simulations, we define the sensing range as a regular polygon, and the sensing radius as the apothem. With the increase of the apothem, a larger area of the disaster map is successfully restored. Fig. \ref{fig:PANDGAMA2} shows that as the unit area content size increases, the effective recovery area degrades since the increase in content size will prolong the transmission time. If the transmission time increases, the probability of successful transmission will decrease. Simultaneously, increasing the transmit power can also effectively enhance the recovery area. \par

Fig. \ref{fig:sanzhong2} shows the effects of the transmission power on the effective recovery area. we compare the proposed SACRL with four benchmarks. Due to the contact time constraint, the probability of a large amount of content being successfully transmitted in a single pass is not as high as the probability of coding into multiple fragments. The effective recovery area expands with higher transmit power, because the higher transmit power increases the number of CUs that can successfully transmit the map and improves the probability of successful transmission. In Fig. \ref{fig:sanzhong1}, the content size per unit area of the disaster map collected by SUs is set to 73-83$Kbits$. When the content is relatively large, the overall effective recovery area will show a downward trend. Since the contact time between the UAV and GV is limited, the content size affects the probability of successful transmission. when the unit area size is 83$Kbits$, the proposed algorithm is 35$\%$ better than NCT algorithm.\par
Fig. \ref{fig:time} shows the relationship between the cumulative effective recovery area and time. Assuming that the time of each disaster map acquisition by the UAV is fixed. The content size per unit area of the disaster map is set to 60$Kbits$. The result indicates that the proposed scheme can continuously and steadily update the disaster map as time grows.
\vspace{-0.1cm}
\section{Conclusions}
\vspace{-0.1cm}
This article presented a hierarchical caching network architecture for emergency communications, which consists of sensing layer, caching layer and transmission layer.  Under the architecture, for typical disaster map transmission applications scenarios, we designed a disaster map collection framework that integrated coded caching technologies. Based on this, we developed a DRL-based algorithm to maximize the probability of successful transmission. The effectiveness of the proposed architecture was demonstrated via detailed simulations highlighting its effectiveness in enhancing map transmission performance under challenging conditions.


\begin{thebibliography}{9}
\vspace{+3pt}
\bibitem{yingji}
L. Wang, J. Zhang, J. Chuan, R. Ma and A. Fei, "Edge Intelligence for Mission Cognitive Wireless Emergency Networks," \emph{IEEE Wireless Communications, }vol. 27, no. 4, pp. 103-109, August. 2020.
\bibitem{livemap}
Q. Wei, L. Wang, L. Xu, Z. Tian and Z. Han, "Hierarchical Coded Caching for Multiscale Content Sharing in Heterogeneous Vehicular Networks," \emph{IEEE Transactions on Vehicular Technology}, vol. 71, no. 6, pp. 5770-5786, June. 2022.
\bibitem{disaster}
M. Deruyck, J. Wyckmans, W. Joseph, and L. Martens,“ Designing UAV-aided emergency networks for large-scale disaster scenarios", \emph{EURASIP Journal on Wireless Communications and Networking}, vol. 79, pp.1-12, Apr. 2018.
\bibitem{1}
F. Ahmad, H. Qiu, R. Eells, F. Bai, G. Motors, and R. Govindan, “{CarMap}: fast 3D feature map updates for automobiles", in \emph{Proc. 17th USENIX Symposium on Networked Systems Design and Implementation (NSDI 20)}, Santa Clara, CA, USA, Feb. 1063-1081, 2020.
\bibitem{jiqunHD}
Y. Wu, Y. Shi, Z. Li and S. Chen, “A cluster-based data offloading strategy for high definition map application," in \emph{Proc. IEEE 91st Vehicular Technology Conference (VTC2020-Spring)},  May. 1-5, 2020.
\bibitem{6--}
J. Pedersen, A. G. i Amat, J. Goseling, F. Brännström, and E. Rosnes, “Dynamic Coded Caching in Wireless Networks," \emph{IEEE Transactions on Communications,} vol. 69, no. 4, pp. 2138-2147, April. 2021.
\bibitem{d2w}
L. Wang, H. Wu, Z. Han, P. Zhang and H. V. Poor, "Multi-Hop Cooperative Caching in Social IoT Using Matching Theory," \emph{IEEE Transactions on Wireless Communications}, vol. 17, no. 4, pp. 2127-2145, April. 2018.
\bibitem{521}
L. Zheng, Q. Chen, Q. Yan and X. Tang, “Decentralized Coded Caching Scheme With Heterogeneous File Sizes," \emph{IEEE Transactions on Vehicular Technology,} vol. 69, no. 1, pp. 818-827, Jan. 2020.
\bibitem{12--}
R. Wang, J. Zhang, S. H. Song and K. B. Letaief, “Mobility-aware caching in D2D networks," \emph{IEEE Transactions on Wireless Communications}, vol. 16, no. 8, pp. 5001-5015, Aug. 2017.
\bibitem{MDS}L. Wang, H. Wu, Y. Ding, W. Chen, and H. V. Poor, “Hypergraph-based wireless distributed storage optimization for cellular D2D underlays," \emph{ IEEE Journal on Selected Areas in Communications}, vol. 34, no. 10, pp. 2650-2666, Oct. 2016.
\bibitem{14--}
M. R. Bhatnagar, “On the capacity of Decode-and-Forward relaying over rician fading channels," \emph{IEEE Communications Letters}, vol. 17, no. 6, pp. 1100-1103, Jun. 2013.
\bibitem{contact time}
L. Wang, H. Tang and M. Čierny, “Device-to-Device Link Admission Policy Based on Social Interaction Information," \emph{IEEE Transactions on Vehicular Technology}, vol. 64, no. 9, pp. 4180-4186, Sept. 2015.
\bibitem{dqn}V. Mnih, K.  Kavukcuoglu, and D. Silver, et al., “Human-level control through deep reinforcement
learning,” \emph{  Nature}, vol. 518, no. 7540, pp. 529–533, Feb. 2015.

\bibitem{drl}H. Ye, G. Y . Li, and B. F. Juang, “Deep reinforcement learning based
resource allocation for V2V communications,”\emph{   IEEE Transactions on
Vehicular Technology}, vol. 68, no. 4, pp. 3163-3173, Apr. 2019.
\bibitem{pso}
H. Wu, F. Lyu, C. Zhou, J. Chen, L. Wang and X. Shen, "Optimal UAV Caching and Trajectory in Aerial-Assisted Vehicular Networks: A Learning-Based Approach," \emph{ IEEE Journal on Selected Areas in Communications,} vol. 38, no. 12, pp. 2783-2797, Dec. 2020.
\bibitem{noncode}
W. Wen, Y. Cui, F. Zheng, S. Jin, and Y. Jiang, “Random Caching Based Cooperative Transmission in Heterogeneous Wireless Networks," \emph{IEEE Transactions on Communications,} vol. 66, no. 7, pp. 2809-2825, July. 2018.
\end{thebibliography}
\end{document}